\documentclass{aa}  

\usepackage{graphicx}
\usepackage[]{natbib}
\usepackage{txfonts}
\usepackage{color}

\begin{document}

        \authorrunning{Ignacio F. Ranea-Sandoval \& Federico Garc\'{\i}a}

   \title{Magnetised accretion discs in Kerr spacetimes}
   \author{Ignacio F. Ranea-Sandoval
          \inst{1}\thanks{Fellow of CONICET, Argentina.}
          \and
          Federico Garc\'{\i}a
          \inst{2,3}$^\star$
          }

\institute{Grupo de Gravitaci\'on, Astrof\'{\i}sica y Cosmolog\'{\i}a, Facultad de Ciencias Astron\'omicas y Geof\'{\i}sicas, Universidad Nacional de La Plata, Paseo del Bosque, B1900FWA La Plata, Argentina. \\
\email{iranea@fcaglp.unlp.edu.ar}
\and
Instituto Argentino de Radioastronom\'{\i}a (CCT La Plata, CONICET), C.C.5, (1894) Villa Elisa, Buenos Aires, Argentina.\\
\email{fgarcia@iar-conicet.gov.ar}
\and
Facultad de Ciencias Astron\'omicas y Geof\'{\i}sicas, Universidad Nacional de La Plata, Paseo del Bosque, B1900FWA La Plata, Argentina.
}

   \date{Received ; accepted }


  \abstract
   {Observational data from \mbox{X-ray} binary systems provide strong evidence of astronomical objects that are too massive and compact to be explained as neutron or hybrid stars. When these systems are in the thermal (\textup{{\em high/soft}}) state, they emit mainly in the ${0{.}1-5}$~keV energy range. This emission can be explained by thin accretion discs that formed around compact objects like black holes. The profile of the fluorescent iron line is useful to obtain insight into the nature of the compact object. General Relativity does not ensure that a black hole must form after the complete gravitational collapse of very massive stars, and other theoretical models such as naked singularities cannot be discarded. The cosmic censorship conjecture was proposed by Penrose to avoid these possibilities and is yet to be proven.}
   {We study the effect caused by external magnetic fields on the observed thermal spectra and iron line profiles of thin accretion discs formed around Kerr black holes and naked singularities. We aim to provide a tool that can be used to estimate the presence of magnetic fields in the neighbourhood of a compact object and to probe the cosmic censorship conjecture in these particular astrophysical environments.}
   {We developed a numerical scheme able to calculate thermal spectra of magnetised Page-Thorne accretion discs formed around rotating black holes and naked singularities as seen by an arbitrary distant observer. We incorporated two different magnetic field configurations: uniform and dipolar, using a perturbative scheme in the coupling constant between matter and magnetic field strength. Under the same assumptions, we obtained observed synthetic line profiles of the $6{.}4$~keV fluorescent iron line.}
   {We show that an external magnetic field produces potentially observable modifications on the thermal energy spectrum and the
fluorescent iron line profile. Thermal energy spectra of naked singularities are harder and brighter than those from black holes, and in addition, peak and cut-off energies are affected by the external magnetic field. Moreover, iron line profiles of slowly rotating black holes suffer more changes by a uniform magnetic field, while nearly extremal black holes and naked singularities are more altered in the dipolar case. Based on our calculations, we discard the possibility of modelling the archetypal black-hole candidate in \mbox{Cygnus X-1} as a naked singularity.}
   {Comparison of our models with observational data can be used to probe the cosmic censorship conjecture and to estimate the existence and global geometry of magnetic fields around compact objects by fitting the thermal energy spectra and iron line profiles of \mbox{X-ray} binaries.}


   \maketitle
%

\section{Introduction}

Answering the question of the final fate of an initial mass distribution after gravitational collapse is an active field of research in relativistic astrophysics. General Relativity (GR) predicts the appearance of spacetime singularities. However, it does not ensure the formation of an event horizon covering them.

In 1969, Roger Penrose proposed the cosmic censorship conjecture (CCC) \citep[see][for a review in the subject]{CCC}. In its weak form, the CCC roughly states that as a consequence of the gravitational collapse of ``normal'' matter, every spacetime singularity should be covered by an event horizon. Determining whether this conjecture is true or not is amongst the most important open problems of GR. Many different lines of thought were pursued to shed some light onto this subject. However, although great effort has been extended over the past 40 years, there is still no definitive answer to whether (weak and strong) CCCs are valid \citep[see][and references therein]{joshi}.

The possibility of turning a black hole into a naked singularity by accretion of different types of particles was analysed in several theoretical researches, and a variety of conclusions have been proposed: from works finding that an extreme black hole cannot accrete particles that would make its horizon disappear to works that contradict this conclusion by analysing accretion of test particles by nearly extreme black holes \citep[see, for example, ][for studies related to Kerr and Reissner-Nordstr\"om solutions]{BHtoNS1,BHtoNS2,BHtoNS3,BHtoNS4,BHtoNS5}. 

Numerical experiments with different types of simple fluid configurations have been performed to finally form a naked singularity \citep[see, for example,][]{num_code1,num_code2,num_code3}, showing, in principle, counterexamples to a weak CCC. These counterexamples are often criticised because of the extremely symmetrical initial configurations and simplifying assumptions needed \citep{CCC}. 

Following a rather different approach, the stability under linear perturbations of the most relevant naked singular spacetimes was systematically studied for the Schwarzschild and Reissner-Nordstr\"om solutions \citep{dottietal11,dottietal12,dottietal13} and for the Kerr spacetime \citep{dottietal21,dottietal23}. The general conclusion of this series of papers is that all naked singularities are unstable against gravitational linear perturbations. Other authors studied the stability of the so-called superspinars and a different family of instabilities were found \citep{sspinar1,sspinar2}.

The non-linear nature of Einstein's equations almost prohibits studying spacetimes of general nature, therefore only highly symmetric spacetimes have been studied from a theoretical point
of view (but this is really adequate to model some aspects of astronomical objects). This is a shortcoming when trying to answer such a complex question as the validity of the CCC. However, this type of studies may become relevant because they serve to compare theoretical predictions with observational data.

Despite the great efforts made, there is no conclusive evidence about the actual nature of ultra-compact objects from the observational point of view, and no direct evidence of an event horizon has been found to date. For this reason, finding observable features that could help us to distinguish between black holes and naked singularities should be considered relevant. Any observation of this distinctive feature would greatly enhance the current black hole paradigm.

In this context, a common feature observed in a wide variety of astrophysical environments, particularly near compact objects, is the formation of accretion discs. A key ingredient for studying realistic accretion-disc models are equatorial circular orbits \citep{sys,pyt}. These equatorial circular orbits were studied in the field of a rotating naked singularity by \citet{Stuchlik,Stuchlik2}. A recent comparative study of properties of accretion discs around black holes and naked singularities can be found in \citet[][and references therein]{kovacs}.

The study of magnetised accretion discs formed around ultra-compact objects is mainly relevant because they are believed to be engines of the not yet completely understood relativistic jets. These extremely energetic phenomena appear on a wide range of scales: from several millions of solar mass active galactic nuclei (quasars) to stellar-mass black holes or neutron stars (micro-quasars). As a possible explanation of jet formation, the so-called disc-jet coupling has been proposed by several authors \citep[see, for example, ][]{d-jet1, d-jet2, d-jet3, d-jet4}. Two of the most widely accepted mechanisms reported to explain the energetics involved in a relativistic jet are based on rotational energy extraction from the central black hole: the Blandford-Znajek process \citep{BZ} and Penrose's mechanism \citep{Pmech, Pmech2a, Pmech2b}. 

Although accretion-disc physics are governed by a complex non-linear combination of different processes such as gravito-magnetohydrodynamics, turbulent viscous fluids, and radiation fields, equatorial circular orbits play a significant role in analytical accretion-disc models. Accretion discs and compact objects are also associated with
observational evidence related to external magnetic fields. Investigating the differences between the effects of simple magnetic field configurations on the circular orbits around a black hole and/or a naked singularity can, in principle, help to distinguish not only the nature of the compact object, but also the strength and geometry of the local magnetic field. 

The effects of magnetic fields on the trajectories of particles orbiting around a rotating black hole and the changes in the innermost stable circular orbit were analysed in \citet{pv, wiita,iyer.et.al}. Changes in the properties of these trajectories may give rise to observable quantities that might allow discerning between different available models for ultra-compact objects. \citet{rsv} calculated the changes in the innermost stable circular orbit and in the marginally bounded orbits using an analytical perturbative approach that allows not only testing previous numerical results, but also reinforcing them. This approach to the problem limits the study of circular trajectories to the plasma case, which is relevant in accretion-disc models. 

In the past years, new techniques based on a spectral analysis of \mbox{X-ray} continuum spectra were developed. These advances allowed estimating not only the mass of ultra-compact objects, but also their spin parameter \citep{obs_a}.
 
Moreover, broad, skewed, and usually double-peaked fluorescent iron $K\alpha$ emission lines have been observed during the past 20 years in a wide variety of galaxies and \mbox{X-ray} binary systems \citep{kalpha}. The first models of line profiles from accretion discs around Kerr black holes were obtained during the 1990s \citep[see, for example, the works by][]{laor,cadez.et.al,fanton.et.al}. In a recent work, \citet{schee.stuchlik} analysed the differences between the profiles of lines generated in the field of rotating black holes and superspinars.

We here used the results presented in \citet{rsv} and an adapted version of the \texttt{Fortran 95} free source code YNOGK \citep{ynogk} to model the observed thermal energy spectra and $K\alpha$ iron line profiles from magnetised accretion discs formed in general Kerr spacetimes, representing black holes and naked singularities. 

Comparisons between observational data from \mbox{X-ray} binary systems and our model can be used as a tool to obtain insight into the physical properties related to the compact object present in these binary systems and also to test the CCC in astrophysical environments. Moreover, because we took into account the effect caused by an external magnetic field on observable quantities, our results can be implemented to estimate the magnetic field strength and geometry in the neighbourhood of the accretion disc.  

Our paper is organised as detailed below. In Sect. \ref{mag-disc} we present the theoretical framework used to construct our accretion-disc model. In Sect. \ref{num-code} we describe our numerical code and how we adapted the public ray-tracing YNOGK code to include Kerr naked singularities and magnetic field effects on the trajectories followed by charged particles. Section \ref{results} is devoted to present in detail our results for observed thermal energy spectra and emission line profiles for accretion discs in a generic Kerr background spacetime with an external (uniform or dipolar) magnetic field. Finally, we present the comparison between our model and observational data of the thermal energy spectra of \mbox{Cygnus X-1}, and discuss our main results in Sect. \ref{conclusions}.

\section{Magnetised accretion discs} \label{mag-disc}

In various astronomical systems, magnetic fields play a decisive role for gas dynamics. Some relevant examples are situations in which the global magnetic field structure governs plasma motion: jet physics, solar coronal activity, and accretion processes into compact objects such as white dwarfs, neutron stars, and completely collapsed objects. Moreover, small-scale magnetic field features might be relevant in the angular momentum transport problem in accretion discs \citep[see, for example,   ][]{somov}.

Accretion processes onto an astrophysical compact object  are extremely complex phenomena that involve understanding a series of different physical theories (General Relativity coupled to magnetohydrodynamics, turbulence, etc.). It is believed that when the accretion process occurs through the slow movement of equatorially concentrated material that falls inwards, an accretion disc truncated at an inner radius is formed. A proper model of where this disc truncates is of central importance in accretion-disc theories.

The pioneering work of \citet{sys} and its relativistic generalisation developed by \citet{pyt} sets the grounds for studying thin, optically thick accretion discs in astrophysical environments.

The structure of the magnetic field in the neighbourhood of the system formed by the compact object plus the accretion disc is extremely complex. In a series of works, \citet{CB-1} and \citet{CB-2} proposed that a mechanism called the Poynting-Robertson cosmic battery (CB) might be responsible for generating an {\it in situ} and maintaining a central dipolar magnetic field. Another viable magnetic field configuration is one that presents almost no change in its magnitude in the equatorial region where the disc is present.

Following these results and those obtained by \citet{pttson}, we here present results related with two different magnetic field configurations: uniform and dipolar. 

Page-Thorne accretion discs are formed by matter orbiting in equatorial circular geodesics. We uses the results for circular trajectories followed by charged particles in a Kerr spacetime with an external global magnetic field presented in \citet{rsv} to model magnetised accretion discs. As most accretion-disc models, ours is not completely self-consistent because the matter in the disc and in the magnetic field does not alter the spacetime geometry. Moreover, the magnetic field structure is not tied to plasma motion, and we assumed that the currents are due to the bulk motion of the fluid in the disk alone. For these reasons, we only treated magnetic fields of small strengths that are weakly coupled to a plasma with a very low conductivity.

After the structure of this disc was obtained, we calculated
the observed thermal energy spectra and fluorescent $K\alpha$ iron line profiles, focusing on the modifications introduced by the external magnetic field.

\subsection{Kerr spacetime}

An object with mass $M$ and angular momentum $J=a M$ can be described, using Boyer-Lindquist coordinates, with the solution reported
by \citet{kerr} to vacuum Einstein field equations. Its line element is expressed as

\begin{eqnarray}
{\rm d}s^2  &=& \frac{\Delta-a^2 \sin ^2 \theta}{\Sigma}{\rm d}t^2 + 4 a M \sin ^2 \theta\frac{r}{\Sigma}{\rm d}t{\rm d}\phi - \frac{\Sigma}{\Delta}{\rm d}r^2 \nonumber \\
&&{} - \Sigma {\rm d}\theta ^2 - 
\left(\frac{(r^2+a^2)^2-\Delta a^2 \sin ^2\theta}{\Sigma}\right)\sin ^2\theta {\rm d}\phi ^2 , \nonumber \label{kerrbl}
\end{eqnarray}

\noindent where $\Delta = r^2-2Mr+a^2$, $\Sigma = r^2+a^2 \cos ^2 \theta$ and the metric signature $+---$ and geometric units $c=G=1$ are used. 

This solution has a physical singularity, the so-called ring singularity, at the points of spacetime in which $\Sigma = 0$ in which the theory breaks down. In the sub-extreme case ($0<a<M$), the spacetime represents a black hole with two horizons located at the zeros of the function $\Delta$. The external horizon causally disconnects the interior from the exterior region, hiding the curvature singularity. In the extreme case ($a=M$), the two horizons collapse into one as the root of $\Delta$ becomes double. In the super-extreme case ($a>M$), the roots of $\Delta$ become complex conjugates and no horizons are present. In this case, the ring singularity is causally connected to future null infinity, and this spacetime represents what is commonly called a rotating (or Kerr) naked singularity.

\subsection{Motion in the equatorial plane}

In the pioneering work of \citet{carter} \citep[see also the recent work by][]{Stuchlik2}, the fundamental aspects of the geodesic motion in Kerr spacetime were coined. The discovery of a Killing tensor allowed separating the Hamilton-Jacobi equation for this spacetime. 

We only work with electromagnetic fields that preserve the stationarity and axial symmetry of the background spacetime. Therefore the Killing nature of the $\partial _{t,\phi}$ vector fields remains
unchanged, and we have that in addition to the particle charge, $e$, and rest mass, the canonical angular momentum, $L$, and energy, $E$, are also conserved along the trajectory. These two quantities can be expressed as

\begin{eqnarray}
L &=& U_\phi + e A_\phi ,\nonumber \\
E &=& -\left( U_t + e A_t \right) \nonumber,
\end{eqnarray}

\noindent where we normalised all quantities with respect to the rest mass of the particle considered. $U_{t,\phi}$ are components of the particle four-velocity, and $A_{t,\phi}$ is the four-vector potential.

We used the arguments presented in \citet{bardeen} and defined an effective potential, $V_{\rm eff}$, for the radial motion. Using the results of \citet{pv}, normalizing all the quantities with respect to $M$ (but keeping the same notation), the effective potential reads

\begin{eqnarray}
V_{\rm eff} = E = -A_t +\frac{2a\left(L-A_\phi\right) + \sqrt{\Delta \left( r^2 \left(L-A_\phi\right)^2 +r R\right)}}{R}, \label{Veff_mag}
\end{eqnarray}

\noindent where $R = r^3+a^2r+2a^2$.

Below we examine each of the two particular magnetic field configurations we plan to study, but first we start by presenting the basics aspects of the equatorial geodesic motion in the non-magnetised scenario.

\subsubsection{Non-magnetised case}

Because of its mathematical simplicity, it is illuminating to review the process of finding the conserved quantities and innermost stable orbit for massive test particles (charged or neutral) following equatorial circular trajectories in non-magnetised Kerr spacetime. The general structure of the reasoning used in this part of the work is used to study the magnetised case we present below.

To analyse this case, we use Eq.~(\ref{Veff_mag}) and set all the components of the four-vector potential to 0. In the framework of the effective potential, equatorial circular orbits are described by 

\begin{eqnarray}
V_{\rm eff}\left(r_c;L,E\right) &=& \frac{1}{2}\left(E^2 - 1\right), \label{eq-circ-a}\\
\left. \frac{{\rm d}V_{\rm eff}}{{\rm d}r}\right|_{r=r_c} &=& 0, \label{eq-circ-b}
\end{eqnarray}

\noindent where $r_c$ is the radius of a generic circular orbit. Solving this pair of equations, we determine the values of the energy, $E$, and angular momentum, $L$, corresponding to a given value of $r_c$.

After substituting $u=1/r$, Eqs.~(\ref{eq-circ-a}) - (\ref{eq-circ-b}) read
\begin{eqnarray}
E^2-1 &=& -2u + \left[L^2 - a^2\left(E^2-1\right)\right] {u^2} - 2\left(L-aE\right)u^3, \nonumber  \\
0 &=& -1 + \left[L^2 - a^2\left(E^2-1\right)\right] u - 3\left(L-aE\right)u^2. \nonumber
\end{eqnarray}

When we combine these two equations, introduce a new conserved quantity $x=L-aE,$ and rearrange the terms properly, we can rewrite these basic equations as

\begin{eqnarray} 
 E^2 &=& \left(1-u\right) + x^2u^3, \label{equsis1a} \\
2xaEu &=& x^2u\left(3u-1\right) - \left(a^2u - 1\right) \label{equsis1b}.
\end{eqnarray}

After eliminating $E,$  we obtain a biquadratic equation that
can be solved for $x^2$ to derive

\begin{eqnarray}
x^2 &=& \frac{\left(a\sqrt{u} \pm 1 \right)^2}{u\left(1-3u\mp2a\sqrt{u^3}\right)}. \label{xsq}
\end{eqnarray}

In the previous expression, the lower (upper) signs correspond to co-rotating (counter-rotating) orbits.

The stability condition for circular orbits requires \mbox{${\rm d}^2V_{\rm eff}/{\rm d}u^2 \ge 0$}. From this inequality it follows that these orbits must have \mbox{$x<0$}, and thus energy, angular momentum, and angular velocity of equatorial stable circular orbits can be derived as

\begin{eqnarray} 
E &=& \frac{1-2u \mp au^{3/2}}{\sqrt{1-3u\mp au^{3/2}}},  \label{cons-E-nm}\\
L &=& \mp \frac{1+a^2u^2\pm 2au^{3/2}}{\sqrt{u}\sqrt{1-3u\mp2au^{3/2}}}, \label{cons-L-nm} \\
\Omega &=& \frac{1}{a \pm u^{-3/2}} \label{cons-O-nm}. 
\end{eqnarray}

\begin{figure}
\centering
\includegraphics[height=0.45\textwidth, angle=-90]{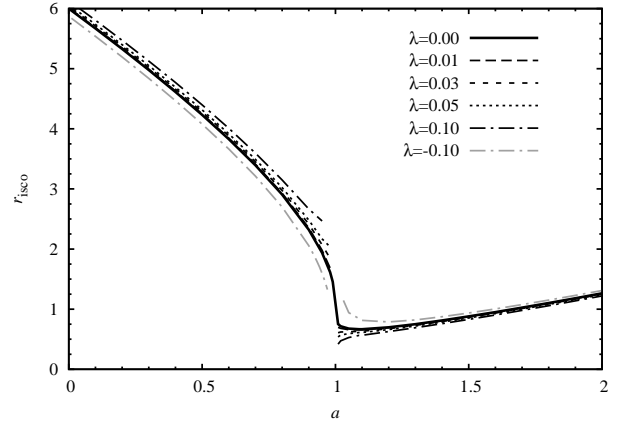}
 \caption{Radii (in gravitational radii, $r_{\rm g} = GM/c^{2}$) of the innermost stable circular orbit, $r_{\rm isco}$, as a function of the spin parameter, $a$. The black thick line corresponds to the non-magnetised case (which, at first order in $\lambda$, remains unchanged for the uniform magnetic field case). Dashed lines correspond to first-order correction to $r_{\rm isco}$ for the dipolar magnetic field configuration for different values of $\lambda$ (see the legend).}
 \label{fig0}
\end{figure}

The marginally or innermost stable circular orbit, $r_{\rm isco}$, can be obtained from Eqs.~(\ref{cons-E-nm})-(\ref{cons-O-nm}) by means of the equality \mbox{${\rm d}^2V_{\rm eff}/{\rm d}u^2 = 0$}, which yields
\begin{equation}
r_{\rm isco} = 3 + Z_2 -\sqrt{\left(3-Z_1\right)\left(3+Z_1+2Z_2\right)}, \label{r0st}
\end{equation}

\noindent where
\begin{eqnarray}
Z_1 &=& 1 + \left(1 - a^2\right)^{1/3}\left[\left(1+a\right)^{1/3} + \left(1-a\right)^{1/3}\right], \nonumber \\
Z_2 &=& \sqrt{3a^2 + Z_1^2}. \nonumber
\end{eqnarray}

In Fig.~\ref{fig0} we plot $r_{\rm isco}$ with a black thick line as a function of the spin parameter $a$. The non-rotating Schwarzschild solution $r_{\rm isco} = 6 r_g$ is recovered for $a=0$.

\subsubsection{Magnetised case}

In presence of an external magnetic field with the characteristics of the fields we are interested in, a similar analysis can be performed.
 
The expressions involved are more complicated and the results are harder to interpret. In the following subsections we study some aspects of this particular problem.

In \citet{wiita} and \citet{iyer.et.al}, the authors analysed the coupling between the effective charge of a volume element and the external magnetic field, quantified by the $\lambda$ parameter. In these works the force on the fluid element due to the magnetic field was calculated using the Lorentz force, where the current is assumed to be in the direction of the bulk motion of the fluid, which is valid for a very low-conductivity plasma. They found that for a fluid disc that is electrically neutral over large scales, this parameter is in the $\lambda \ll 1$ regime, while for charged test particles, $\lambda$ can easily exceed unity. Although the matter of the disc should be neutral as a whole, some charge separation is expected as a consequence of complex effects related to plasma dynamics. In the weak-coupling limit, where the global dynamics of fluid elements are governed by gravity, the problem can be approached based on the study of the small departures from geodesic paths followed by macroscopically neutral fluid volume elements. Based on these results, we conclude that the perturbative analysis we performed here to calculate circular trajectories can be used to study a fluid relativistic accretion disc.

The procedure for finding the radius, energy, and angular momentum of these trajectories is, essentially, the same as in the non-magnetised case. The main difference are the much more involved mathematical formulae.

To recover the expression of \citet{Hobson} for the non-magnetised case, we worked with an alternative form of the effective potential that we named $U_{\rm eff}$. This expression can be inserted as a function of the $u$-radial coordinate in this way:

\begin{eqnarray}
  U_{\rm eff} &=& U_0 + \lambda U_1 + \lambda ^2 U_2, \label{Ueff}
\end{eqnarray} 

\noindent where $\lambda$ measures, as we have said, the coupling between the external magnetic field and the particles effective charge and the quantities $U_{\rm i}$ are defined as

\begin{eqnarray}
U_0 &=&- {x}^{2}{u}^{3}+ \left( \frac{{x}^{2}}{2}+xEa+\frac{{a}^{2}}{2} \right) {u}^{2}-u \nonumber \\
U_1 &=& \left(Ea-x\right) \left( \frac{{a}^{2}}{2} {u}^{2} - u   + \frac{1}{2} \right)  \label{def2} \\
U_2 &=& \frac{1}{8}\left({u^{-2}}-2{u^{-1}} - 2{{a}^{2}} + 8{a}^{2}u -\left( 3{a}^{2}+4 \right){{a}^{2}u^2}  + 2{a^4u^3} \right). \nonumber
\end{eqnarray}

After imposing the circular orbit condition, the previous equation and its derivative with respect to the $u$-radial coordinate can, in principle, be used to obtain $E$ and $x$ as a function of $u$, similar to Eqs.~(\ref{equsis1a})-(\ref{equsis1b}) for the non-magnetised case. Equation ${\rm d}U_{\rm eff}/{\rm d}u = 0$ is a linear equation for $E,$ and after solving it and inserting the result in an appropriate combination of Eq.~(\ref{Ueff}) and its derivative with respect to $u$, a complete quartic equation for $x$ is derived. This equation is typically solved numerically, but \citet{rsv} adopted a different approach. Based on the exact solution for the non-magnetised case, we treated the magnetised case perturbatively in the $\lambda$ parameter using the following procedure: we expanded the conserved quantities and their equations in a Taylor polynomial and used a few steps of the Newton method to find roots of a transcendental equation to derive an exact solution (up to the perturbation order we worked with).

We constrain our study to astrophysically relevant co-rotating orbits. Using the exact solutions for the non-magnetised case as seeds for the magnetised case, we improved them by means of Newton's method to find analytical solutions to a transcendental equation. After this we obtained the perturbative first-order solutions. These new conserved quantities fully characterise the circular orbits, which are key features for studying accretion-disc models. In the same way, using the first-order conserved quantities and Eq.~(\ref{r0st}) for the $r_{\rm isco}$ in the non-magnetised case as a seed, we obtained the first-order correction to its position.  

To obtain the position of the innermost stable circular orbit, the equation to be solved is

\begin{equation}
\frac{{\rm d}^2 U_{\rm eff}(u,E,x;a,\lambda)}{{\rm d}u^2} = 0.
\end{equation}

Because we consider the weakly coupled case, we can expand in Taylor series the previous equation to obtain

\begin{eqnarray}
\frac{{\rm d}^2 U_{\rm eff}(u,E,x;a,\lambda)}{{\rm d}u^2} &=& \frac{{\rm d}^2 U_{\rm eff}(u^0_{\rm isco},E^0,x^0;a,\lambda = 0)}{{\rm d}u^2} + \nonumber \\
\epsilon_{\rm isco} \left.\frac{{\rm d}^3 U_{\rm eff}}{{\rm d}u^3}\right|_{\lambda=0} &+& \lambda \frac{{\rm d}}{{\rm d}\lambda}\left.\left(\frac{{\rm d}^2U_{\rm eff}}{{\rm d}u^2}\right)\right|_{\lambda = 0} = 0. 
\end{eqnarray}

In this step, we explicitly used the analytical expression obtained for both $E$ and $x$ to the perturbative order we work with. The solution to this equation gives the position of the innermost circular stable orbit for the pair $(a,\lambda)$. It is easy to derive that the first-order correction, $\epsilon_{\rm isco}$, can be written as
\begin{equation}
\epsilon_{\rm isco} = - \lambda \frac{\frac{{\rm d}}{{\rm d}\lambda}\left.\left(\frac{{\rm d}^2U_{\rm eff}}{{\rm d}u^2}\right)\right|_{\lambda = 0}}{\left.\frac{{\rm d}^3U_{\rm eff}}{{\rm d}u^3}\right|_{\lambda=0}}.
\end{equation}

It is important to note that with our perturbative approach we were able to reproduce numerical results obtained in previous works for black holes with low values of the $\lambda$ parameter for a wide range of the spin parameter $a$. As an example, for $a=0{.}5$ and $\lambda =0{.}455$ the value for the co-rotating innermost stable orbit found in \citet{wiita} is $r_{\rm isco} = {2{.}77}$, which is exactly the same value as we found and shows the excellent agreement between the two approaches. In the uniform case, this first correction is zero and the first not null correction is $\mathcal{O}(\lambda ^2)$ (see Eq. (\ref{euisco})).

\subsection{Uniform magnetic field}

The first exact solution for an external electromagnetic field in a Kerr background was found by \citet{WaldB}. He derived the electromagnetic field of a rotating black hole placed in a magnetic field that originally is uniform and aligned with the spin axis;
this configuration can be used as the first rough approximation to the magnetic field produced by the disc in the equatorial plane.

Using Teukolsky's formalism \citep{teuk01}, \citet{pttson} derived explicit expressions for general stationary axially symmetric electromagnetic fields in a Kerr background. Using the explicit solution of the spacetime background given in \citet{pttson}, the four-vector potential can be expressed as
\begin{eqnarray}
A_{\mu} &=& \left(A_t,0,0,A_\phi\right), \nonumber
\end{eqnarray}

\noindent where
\begin{eqnarray}
A_t &=& -aB\left(1-\frac{Mr}{\Sigma}\left(2- \sin ^2 \theta\right)\right), \nonumber \\
A_\phi &=& \frac{B \sin ^2 \theta}{2\Sigma} \left[\left(r^2+a^2\right)^2 - \Delta a^2  \sin ^2 \theta -4Ma^2r\right], \nonumber
\end{eqnarray}

\noindent where $B$ is the magnetic field strength.

The expressions that appear in \citet{pv} can be used without changes to study the effects on the orbits around a naked singularity. Instead of the usual numeric treatment of the problem, we are interested in obtaining analytical expressions in a perturbative way. For this magnetic field configuration, the coupling constant between the effective charge and the magnetic field strength is

\begin{equation}
\lambda _{\rm U} = \frac{e}{m} B M,
\end{equation}

\noindent where $e$ and $m$ are the charge and rest mass of the particles, respectively.

\subsubsection{Innermost stable circular orbit}

For sake of simplicity, it is useful to define a new radial coordinate $u^0_{\rm isco} = 1/r^0_{\rm isco}$. For this magnetic field geometry the linear correction in $\lambda _{\rm U}$ for $r_{\rm isco}$ is null because \mbox{$\frac{{\rm d} E}{{\rm d}\lambda}\left(u^0_{\rm isco},x^0;a,\lambda_{\rm U} =0\right) = 0$}, and thus the first non-zero correction to the $u$-position of the innermost stable circular orbit, $\epsilon ^{\rm U}_{\rm isco}$, due to a uniform magnetic field, is quadratic in $\lambda _{\rm U}$ and its expression is given by

\begin{equation}
\epsilon ^{\rm U}_{\rm isco} = -\frac{1}{2}\lambda _{\rm U}^2\frac{\frac{{\rm d}^2 E}{{\rm d}\lambda ^2}\left(u^0_{\rm isco},x^0;a,\lambda _{\rm U} =0\right)}{\frac{{\rm d} E}{{\rm d}\epsilon ^{\rm U} _{\rm isco}}\bigg|_{\epsilon ^{\rm U} _{\rm isco}=0}\left(u^0_{\rm isco},x^0;a,\lambda _{\rm U} =0\right)}.
\label{euisco}
\end{equation}

Because here we work only with first-order corrections in $\lambda$, for the uniform magnetic field case, we set $r_{\rm isco}=r^0_{\rm isco}$. We refer to \citet{rsv}, where more figures are available.

\subsection{Dipolar magnetic field} \label{Amu.dip}

A possible explanation of how a black hole plus accretion-disc system can maintain a global dipolar magnetic field configuration was proposed in \citet{CB-1,CB-2}. This makes studying this type of magnetic field geometry particularly interesting.

Following the results obtained in \citet{pttson} for a dipolar magnetic field without electrostatic charge, we can write the non-zero components of the four-vector potential as

\begin{eqnarray}
A_t &=& -\frac{3a\mu}{2 \left(1-a^2\right) \Sigma}\left(f_1\left(r,\theta\right) \frac{\ln \Theta}{2 \sqrt{1-a^2}}
-\left(r-M \cos ^2 \theta \right)\right), \nonumber \\
A_\phi &=& -\frac{3\mu \sin ^2 \theta}{4 \left(1-a^2\right) \Sigma} \left(f_2\left(r,\theta\right) +f_3\left(r,\theta\right)\frac{\ln \Theta }{2 \sqrt{1-a^2}} \right), \nonumber
\end{eqnarray}

\noindent where $\mu$ is the dipole moment, which we here consider parallel to the spin axis, $f_i\left(r,\theta\right), i=1,2,3$ are functions defined as

\begin{eqnarray}
f_1\left(r,\theta\right) &=& r\left(r-M\right)+\left(a^2-Mr\right)\cos ^2 \theta ,\nonumber \\
f_2\left(r,\theta\right) &=& \left(r-M\right)a^2\cos ^2 \theta + r\left(r^2+Mr+2a^2\right) , \nonumber \\
f_3\left(r,\theta\right) &=& - \left[r\left(r^3+ a^2r-2Ma^2\right)+\Delta a^2 \cos ^2 \theta \right] \nonumber 
\end{eqnarray}

\noindent and $\Theta$ is given by

\begin{eqnarray}
\Theta &=& \frac{r-M+\sqrt{1-a^2}}{r-M-\sqrt{1-a^2}}. \nonumber
\end{eqnarray}

An important feature of the four-vector potential is that like the background spacetime, it is singular at the ring singularity. The analysis reported by \citet{pv} can be extended for the extreme ($a=M$) Kerr case by studying the particular limiting case and the super-extreme ($a>M$) Kerr case through the analytical extension in the complex plane of the logarithmic function.

For this magnetic field configuration the coupling constant between the effective charge and magnetic field strength is

\begin{equation}
\lambda _{\rm D} = \frac{e \mu}{M^2}.
\end{equation}

\subsubsection{Innermost stable circular orbit}

To study the marginal orbits we adopted the same procedure as in the previous cases and note that contrary to what happened for the uniform magnetic field, the correction to the position of the marginally stable orbit is linear in the parameter $\lambda _{\rm D}$. In Fig.~\ref{fig0} we plot our results for $r_{\rm isco}$ as a function of the spin parameter, $a$, for a set of values of the $\lambda$ parameter. As a black solid line we plot the non-magnetised $r_{isco}$, which is coincident with the uniform magnetic field case at first order in $\lambda$. Because our perturbative approach is invalid in the close neighbourhood of $a=1$, the dashed curves for the magnetised case are not continuous. In this limit, the leading term of the correction is proportional to $\frac{\lambda}{1-a}r_{\rm isco}^0$. This can be used to restrict the range of spin parametrs that can be treated with our perturbative approach in the following manner: If a change of $\delta _{\rm isco} \%$ in the non-magnetised value for $r_{\rm isco}$ is accepted as a perturbation, then for a given value of $\lambda$, our method is invalid in the range $|\lambda|< \delta _{\rm isco} |1-a|$. From the same figure, it follows that for black holes the $r_{\rm isco}$ for the magnetised case when $\lambda >0$ (black lines) is larger than the non-magnetised case one, while for naked singularities, the opposite occurs. For $\lambda <0$ (grey lines), it behaves in the opposite way, as expected.

\section{Numerical code} \label{num-code}

In this section we present the numerical code we developed and
describe how we obtained the observed thermal energy spectra and emission line profiles of accretion discs by solving the
equations presented in the previous section.

To obtain the observed thermal energy spectrum or line profiles, our numerical code proceeds following four steps:

\begin{itemize}
\item We obtain the properties of circular trajectories followed by matter in the disc at radii $r$.
\item We obtain the specific intensity $I(\nu)$ emitted at each radii $r$, on the disc. 
\item We evaluate the relativistic effects suffered by photons emitted on disc as they travel to the observer.
\item We calculate the absorption of photons by the interstellar medium.
\end{itemize}

First of all, from Eqs.~(\ref{cons-E-nm}), (\ref{cons-L-nm}) and (\ref{cons-O-nm}) of Sect. \ref{mag-disc} we obtained the energy, $E$, angular momentum, $L$, and angular velocity, $\Omega$, of circular orbits followed by massive particles, as a function of the radial coordinate, $r$ and the spin frequency of the compact object, $a$, for the non-magnetised case. By a similar procedure, we obtained these quantities from equivalent equations for the uniform and dipolar cases in terms of the coupling constant parameter, $\lambda$. From Eq.~(\ref{r0st}) we calculated the size of the innermost stable circular orbit, $r_{\rm isco}$ for each value of $a$, which we assumed as the inner radius of the discs for non-magnetised and uniform magnetic field cases (as in this case the first-order correction is quadratic in $\lambda$). In a similar manner, we calculated $r_{\rm isco}$ for the dipolar case as a function of $a$ and $\lambda$. At this point, following \citet{11risco}, we set the outer radius of the disc $r_{\rm out} = 11 r_{\rm isco}$.

Following \citet{pyt}, the energy flux as a function of the radial coordinate, $Q(r)$, for a relativistic accretion disc is given by

\begin{equation} \label{Qr}
Q(r) = - \frac{\dot{M}}{4\pi \sqrt{-g_M}}\frac{\Omega _{,r}}{\left(E - \Omega L\right)^2} \int _{r_{\rm isco}} ^{r_{\rm out}} \left( E - \Omega L \right) L_{,r^\prime} {\rm d}r^\prime, 
\end{equation}

\noindent where $\dot{M}$ is the mass accretion rate, which is a free parameter in our model, $g_M$ is the determinant of the metric tensor, and $\Omega_{,r}$ and $L_{,r}$ are the derivatives of the angular velocity and angular momentum with respect to the radial coordinate, for which we deduced analytical expressions for each magnetic field configuration based on Eqs.~(\ref{cons-E-nm}), (\ref{cons-L-nm}) and (\ref{cons-O-nm}) of Sect. \ref{mag-disc}.

Now, using Eq.~(\ref{Qr}) and following the standard method for a steady-state optically thick and geometrically thin accretion disc \citep{pyt,Li.et.al-2005}, the energy flux emitted at the surface of the disc as seen by a locally co-rotating observer, $f(r)$, can be obtained. Then, by means of the Stefan-Boltzmann law for blackbody radiation, an effective temperature, $T_{\rm eff}(r)$, can be deduced as $T_{\rm eff} = \sqrt[4]{Q(r)/ \sigma_{\rm SB}}$, where $\sigma_{\rm SB}$ is the Stefan-Boltzmann constant. Using this temperature distribution on the disc, and assuming thermal blackbody radiation, the intensity emitted at frequency $\nu$, $I(\nu)$ follows the Planck law:

\begin{equation}
I(\nu,r)=\frac{h\nu^3}{\exp\left(\frac{h\nu}{k_{\rm B} T_{\rm eff}(r)}\right)-1}.
\end{equation}

The effective temperature can be corrected using a Comptonisation parameter, $f_{\rm col}$, which takes into account non-thermal effects that might occur in the inner regions of the accretion disc. 

For the line profiles, we assumed monochromatic emission at $\nu_{\rm 0}$ (which, for example, can be set to $6.4$~keV for the $K\alpha$ iron line), represented by a Dirac delta function. Then, assuming a power-law emissivity distribution of index $0<p<4$ \citep[see, for example, ][]{index}, the intensity emitted at the surface of the disc, in the local rest frame, takes the form

\begin{equation}
I\left(\nu,r\right) = \frac{\epsilon_{\rm 0}}{r^{p}} \delta \left(\nu - \nu _{\rm 0}\right),
\end{equation}

\noindent where $\epsilon_{\rm 0}$ is a constant. To obtain the spectrum of the thermal emission and/or the emission line profile, the intensity needs to be integrated over the whole disc. However, because we are interested in the spectra and line profiles as seen by a distant observer at arbitrary position, we incorporated a ray-tracing technique to evaluate the geodesics of photons between a plate placed at the observer's position and the surface of the disc. For this we adapted the public \texttt{Fortran 95} code YNOGK \citep{ynogk}, which includes all the relativistic effects suffered by photons: Doppler boosting, gravitational redshift, and gravitational light bending. We modified the expressions for the conserved quantities of massive particles following circular geodesics to the corresponding expressions described in Sect. \ref{mag-disc} to include the magnetic field effects. We also broke the restriction $|a| \le 1$, which allowed us to study Kerr naked singularities. 

Since $I(\nu)/ \nu ^3$ is invariant along a geodesic \citep{MTW}, the integral on the surface of the disc is converted to an integral over the observer's plate in $\alpha$ and $\beta$ coordinates:

\begin{equation}
F_{\rm obs}\left(\nu_{\rm obs}\right)=\int_{\rm plate} \frac{h\nu_{\rm obs}^3}{\exp\left(\frac{h\nu_{\rm obs}}{g k_{\rm B} f_{\rm col} T_{\rm eff}}\right)-1}{\rm d}\alpha {\rm d}\beta,
\end{equation}

\noindent where $g=\nu_{\rm obs}/\nu_{\rm 0}$ is the redshift of the photon, which incorporates all the relativistic effects already described. Dividing by the energy, the photon number density is $N_{\rm obs}=F\left(\nu_{\rm obs}\right)/E_{\rm obs}$. In the same manner, the line profile results

\begin{equation}
F_{\rm obs}\left(\nu_{\rm obs}\right) = \int_{\rm plate} \frac{\epsilon _{\rm 0}}{r^{p}} g^4 \delta \left(h \left(\nu_{\rm obs} - g \nu_{\rm 0}\right)\right) {\rm d}\alpha {\rm d}\beta.
\end{equation}

Finally,  to fit observational data with the spectra obtained from our model, we used the analytical expression for the interstellar absorption cross-section of \mbox{X-ray} photons given by \citet{int-absX}, parameterised by the hydrogen column density, $N_{\rm H}$.

\section{Results} \label{results}

After presenting the central aspects of our model and numerical code, below we focus on our results and their implications. We divide them into two subsections: one for the thermal energy spectra and the other for the emission line profiles.

\subsection{Thermal energy spectra of magnetised accretion discs}

\begin{figure}
\centering 
\includegraphics[height=0.9\columnwidth,angle=-90]{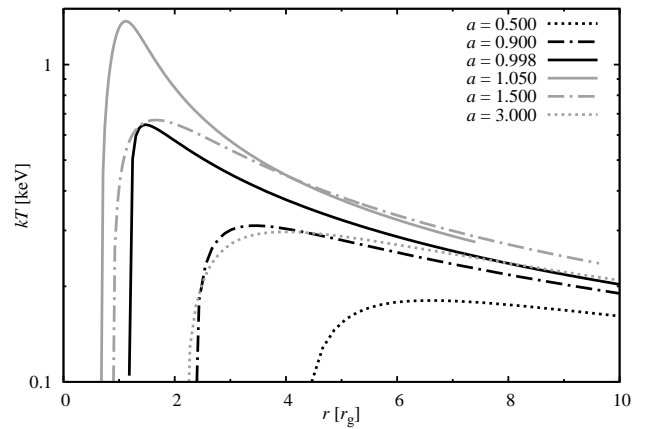}
  \caption{Typical on-disc temperature profiles as a function of the gravitational radius, $r$, in gravitational radii, $r_{\rm g} = GM/c^{2}$, for different values of the spin parameter, $a$. Continuous (dot-dashed) lines correspond to accretion discs formed around black holes (naked singularities). Profiles extend from $r_{\rm in} = r_{\rm isco}$ to $r_{\rm out} = 11 r_{\rm isco}$.}
  \label{plot_T}
\end{figure}

In Fig.~\ref{plot_T} we show on-disc temperature profiles as a function of the radius, $r$, for different values of the spin parameter, $a$. In black we plot profiles for black holes with $a=0.5$, $0.9$, $0.998$ and in grey we plot profiles for naked singularities $a=1.05$, $1.50$, $3.00$. For black holes, the
peak temperature grows in the inner region of the discs for increasing $a$ as $r_{\rm isco}$ decreases. For naked singularities, $r_{\rm isco}$ has the same behaviour until $a=1.089$ and then becomes a growing function with respect to $a$, and thus, the peak temperature decreases for $a \gg 1$. It is noticeable that even for black holes and naked singularities with the same $r_{\rm isco}$, the on-disc temperature profiles look different, which will produce differences in their observed thermal spectra.

\begin{figure}
\centering
\includegraphics[height=0.9\columnwidth, angle=-90]{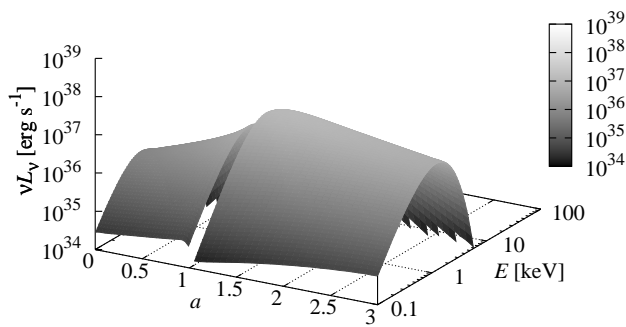}
      \caption{Family of thermal energy spectra as a function of the spin parameter, ${a}$, for $i=30\degr$, ${f_{\rm col} = 1.6}$ and the mass and accretion rate of \mbox{Cygnus X-1}, for a uniform magnetic field with $\lambda=0.1$. The grey scale represents ${\nu L_\nu}$ in erg~s$^{-1}$ (see the legend). Note the sharp change in the spectra around the interface $a \sim 1$.}
         \label{familias}
   \end{figure}

In Fig.~\ref{familias} we plot a family of observed thermal energy spectra obtained for a series of values of the spin parameter $0<a<3$ for magnetised accretion discs with $i=30\degr$, $f_{\rm col}=1.6$, and considering the mass and accretion rate of \mbox{Cygnus X-1} \citep{cygX1}. As a direct consequence of the results described in Fig.~\ref{plot_T} for the on-disc temperature profiles, observed thermal spectra for black holes become harder and more luminous as $a \rightarrow 1$. This behaviour continues for naked singularities as $a<1.089$ and the strongest rising in the brightness occurs in the neighbourhood of $a=1$. Then, for $a>1.089$, the thermal spectra become dimmer and the peak energy decreases, resulting in softer spectra. Two main observables arise from this analysis: peak temperature and cut-off energy, which can be used to compare spectra from different models with observational data.

To analyse the imprints of relativistic effects on the observed thermal energy spectra, in Fig.~\ref{dos-discos} we plot the images that accretion discs would produce in a hypothetical plate where the observer is situated. In grey scale we represent the temperature scale $g T_{\rm col}= g f_{\rm col} T_{\rm eff}$, where $T_{\rm eff}$ is the on-disc temperature and the factor $g$ is a combination of the gravitational redshift and Doppler boosting. Moreover, light-bending effects are clear by the observed shape of the discs, which, at the local rest of frame, are circular. In the left (right) panel of Fig.~\ref{dos-discos} we plot the image of a magnetised black hole (naked singularity) accretion disc with $a=0.99$ ($a=1.05$), $f_{\rm col}=1.6$ and $i=60\degr$. As in the temperature profiles, harder emission comes from the inner regions of the disc, while in the outer region the emission looks softer. The horizontal asymmetry observed in the disc emission is a direct consequence of the Doppler boosting of photons coming from the disc, while the vertical asymmetry in the disc shape is caused by light-bending in the Kerr background spacetime. Because there is no event horizon, a second partial image of the disc is observed inside the inner radius of the main image for the naked singularity. This fact will play an important role for the shape of the emission line profiles.

Now we focus on the effects that the introduction of an external uniform or dipolar magnetic field have on the observed thermal energy spectra. For this, we present in Fig.~\ref{plot6} thermal energy spectra for two pairs of black holes (black lines) and naked singularities (grey lines) for different inclination angles considering non-magnetised ($\lambda=0$) and uniform ($\lambda_{\rm U}=0.1$) and dipolar ($\lambda_{\rm D}=0.1$) magnetised accretion discs.

In the top panels of Fig.~\ref{plot6} we plot thermal spectra for a black hole of $a=0.5$ and a naked singularity of $a=1.5$. For these values of $a$, the $r_{\rm isco}$ is far from the critical value for $a \sim 1$, and thus the accretion discs extend to high $r$ values. Because thermal spectra for dipolar magnetic fields are almost indistinguishable than for non-magnetised discs, the differences are stronger for
uniform fields because they dominate the circular trajectories in the outer regions, where dipolar magnetic fields dilute. In particular, for $\lambda_{\rm U}=0.1$, the peak intensity dims, while the cut-off energy remains insensitive. As expected, for almost extremal black holes and/or naked singularities ($a \sim 1$), in contrast, the effects caused by dipolar magnetic fields are slightly stronger than those from uniform magnetic fields (see bottom panel of Fig.~\ref{plot6}). In both cases, the peak intensity and cut-off energy are affected in similar ways.

\subsection{Emission line profiles}

We continue the presentation of our results with the observed relativistic line profiles. Although our results for line profiles are completely general for any line emitted on the surface of the accretion disc, they become particularly interesting from an astrophysical point of view for the fluorescent $6.4$~keV $K\alpha$ iron line. The profile of the iron line has been studied as a tool for obtaining physical parameters of the accretion discs independently of the thermal energy spectrum \citep{kalpha}. In the same way as for thermal energy spectra, the observed profiles depend on the spin parameter, $a$, the inclination angle, $i$, the geometry of the magnetic field, and the coupling between its strength and the effective charge of the particles of the accretion disc, $\lambda$. Moreover, as the redshift $g$ of photons arriving at the observer's plate depend on Doppler boosting, which, as we already discussed for thermal spectra, is enhanced in the inner region of the disc with respect to the outer region, observed line profiles also depend on the emission law assumed for the radiating disc. In our case, we assumed the standard power-law form $\epsilon_{\rm 0}/r^p$, where $p$ is the emissivity index \citep[see, for instance, ][]{fanton.et.al}. For the same reason, the profiles also depend on the extension of the radiating disc. As for thermal spectra, in this work we assumed $r_{\rm out}=11 r_{\rm isco}$. 

\begin{figure*}
\centering   
\includegraphics[height=0.4\textwidth,angle= -90]{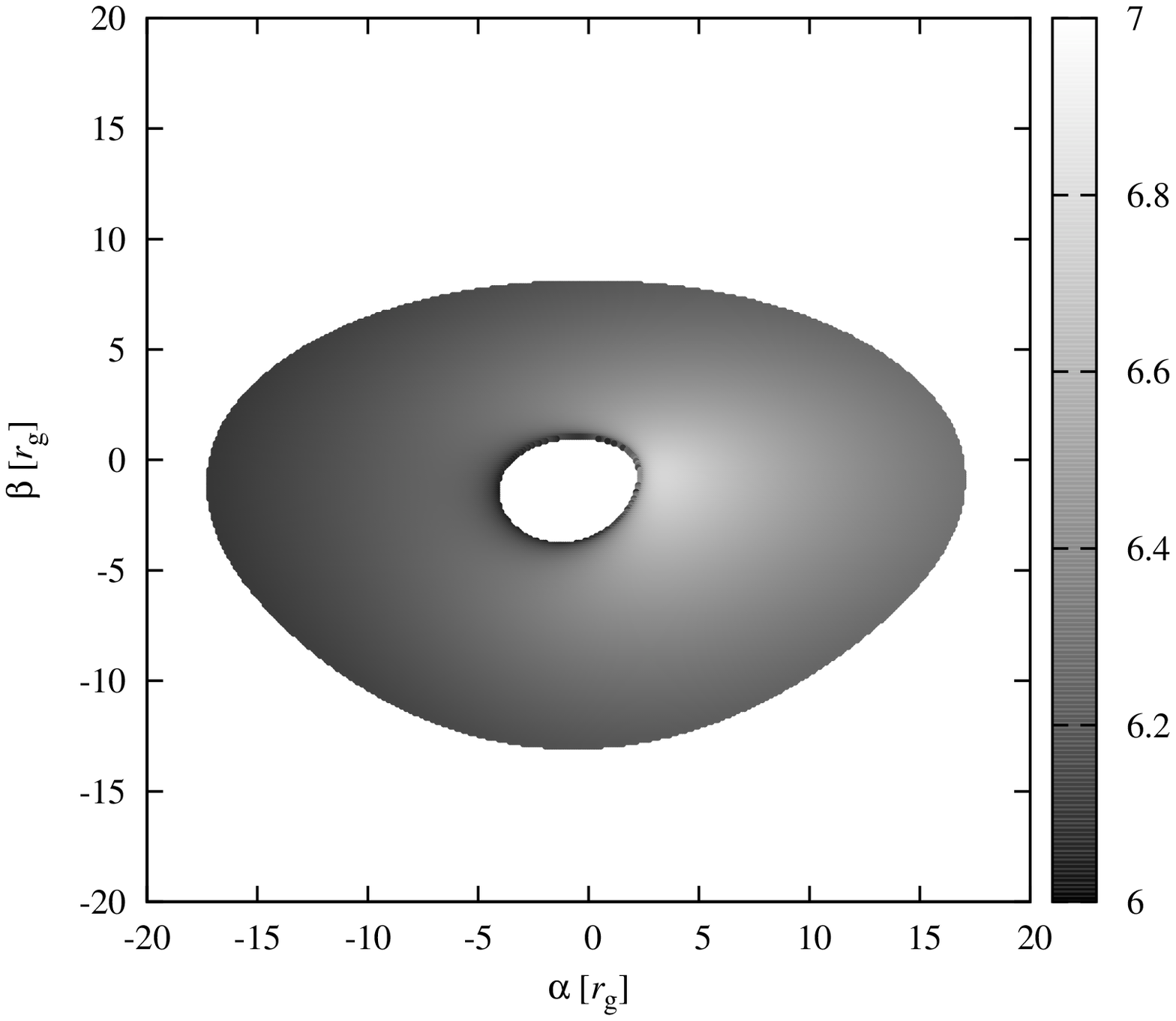} \hspace{1cm} \includegraphics[height=0.4\textwidth,angle= -90]{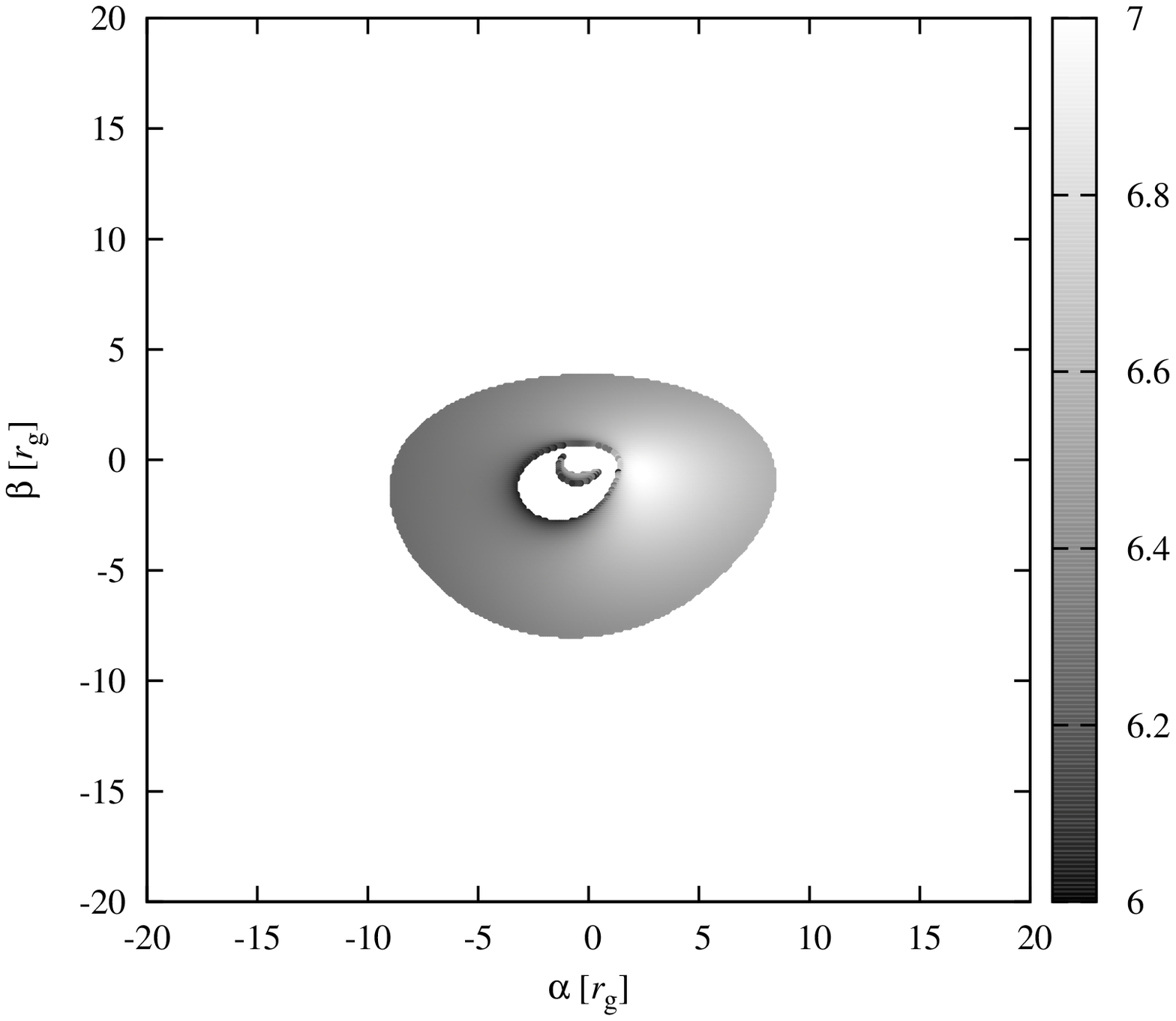}
\caption{Images at the observer's plate of accretion discs tilted with $i=60\degr$, $f_{\rm col}=1.6$, mass and accretion rate of \mbox{Cygnus X-1} and a uniform magnetic field with ${\lambda = 0.1}$. \textbf{Left} panel corresponds to the silhouette of a disc formed around a black hole with $a=0.99$. \textbf{Right} panel corresponds to a disc around a naked singularity with $a=1.05$. The grey scale represents the logarithm of the temperature scale, ${gT_{\rm col}}$, in Kelvin. Cartesian axes represent the impact parameters on the observer's plate in gravitational radii, $r_{\rm g} = {GM/c^2}$.}
\label{dos-discos}
\end{figure*}

\begin{figure*}
\centering
  \includegraphics[height=0.9\textwidth, angle=-90]{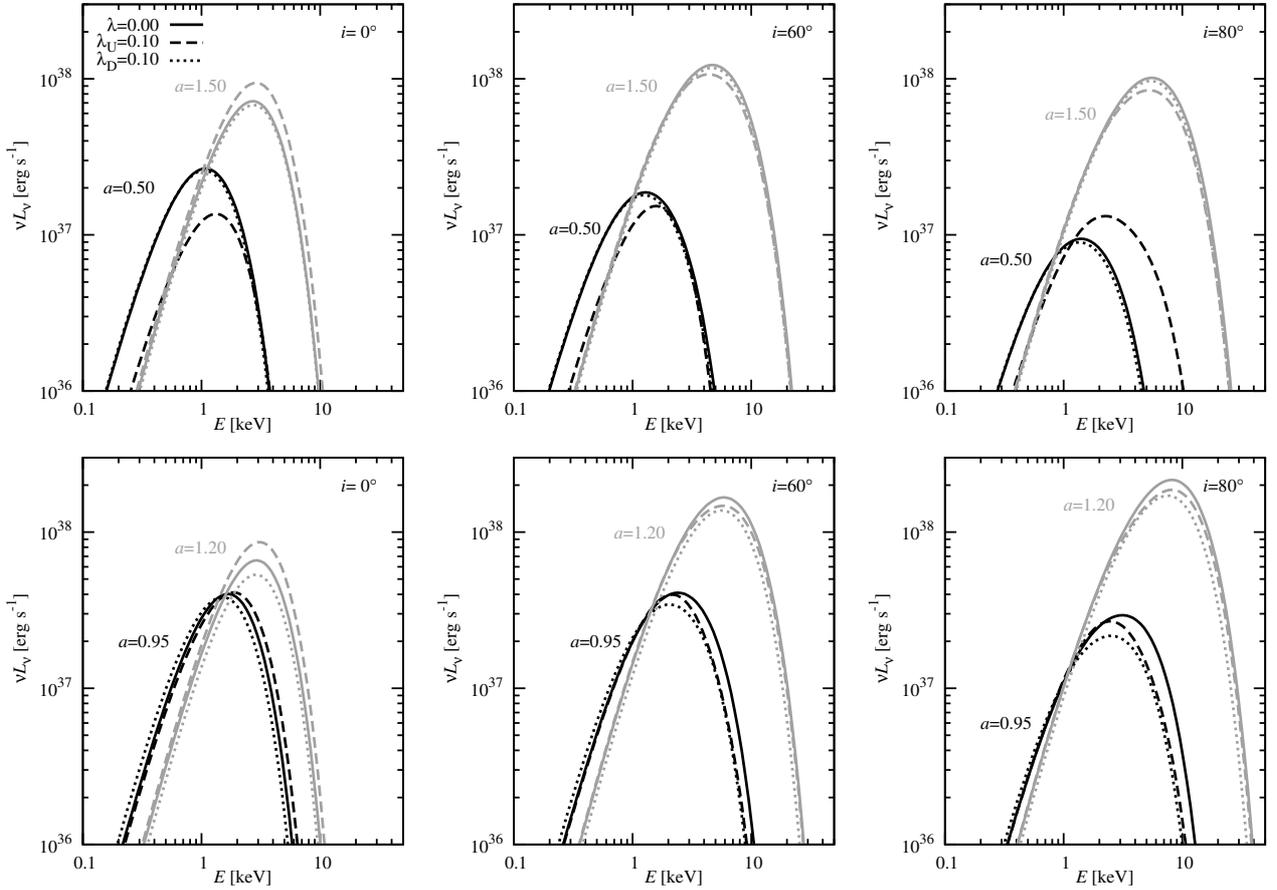}
  \caption{Thermal energy spectra of accretion discs formed around black holes (black lines) and naked singularities (grey lines) for inclinations, from left to right, $i=0\degr$, $60\degr$, $80\degr$. \textbf{Top (bottom)} panels correspond to black holes with $a=0.5$ ($a=0.95$) and naked singularities with $a=1.5$ ($a=1.2$). Solid lines represent the non-magnetised case. Dashed (dotted) lines represent the uniform (dipolar) magnetic field configuration with $\lambda=0.1$.}
  \label{plot6}
\end{figure*}

\begin{figure*}
\centering
  \includegraphics[height=0.9\textwidth, angle=-90]{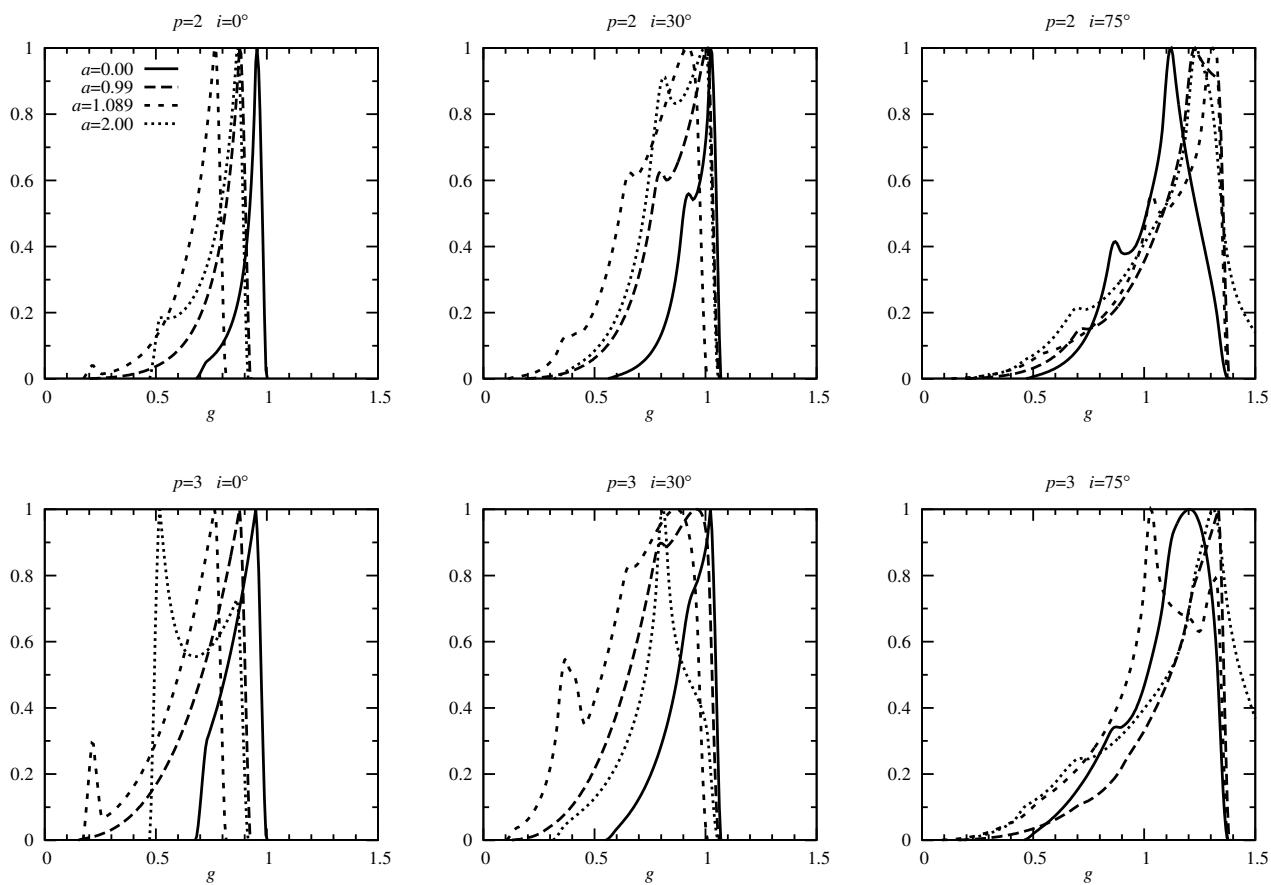}
  \caption{Emission line profiles of non-magnetised accretion discs formed around black holes and naked singularities for four different values of the spin parameter, $a$ (see the legend), for inclinations, from left to right, $i=0\degr$, $30\degr$, $75\degr$. \textbf{Top (bottom)} panels correspond to an emissivity index $p=2$ ($p=3$).}
  \label{plot1}
\end{figure*}

\begin{figure*}
\centering
  \includegraphics[height=0.9\textwidth,angle=-90]{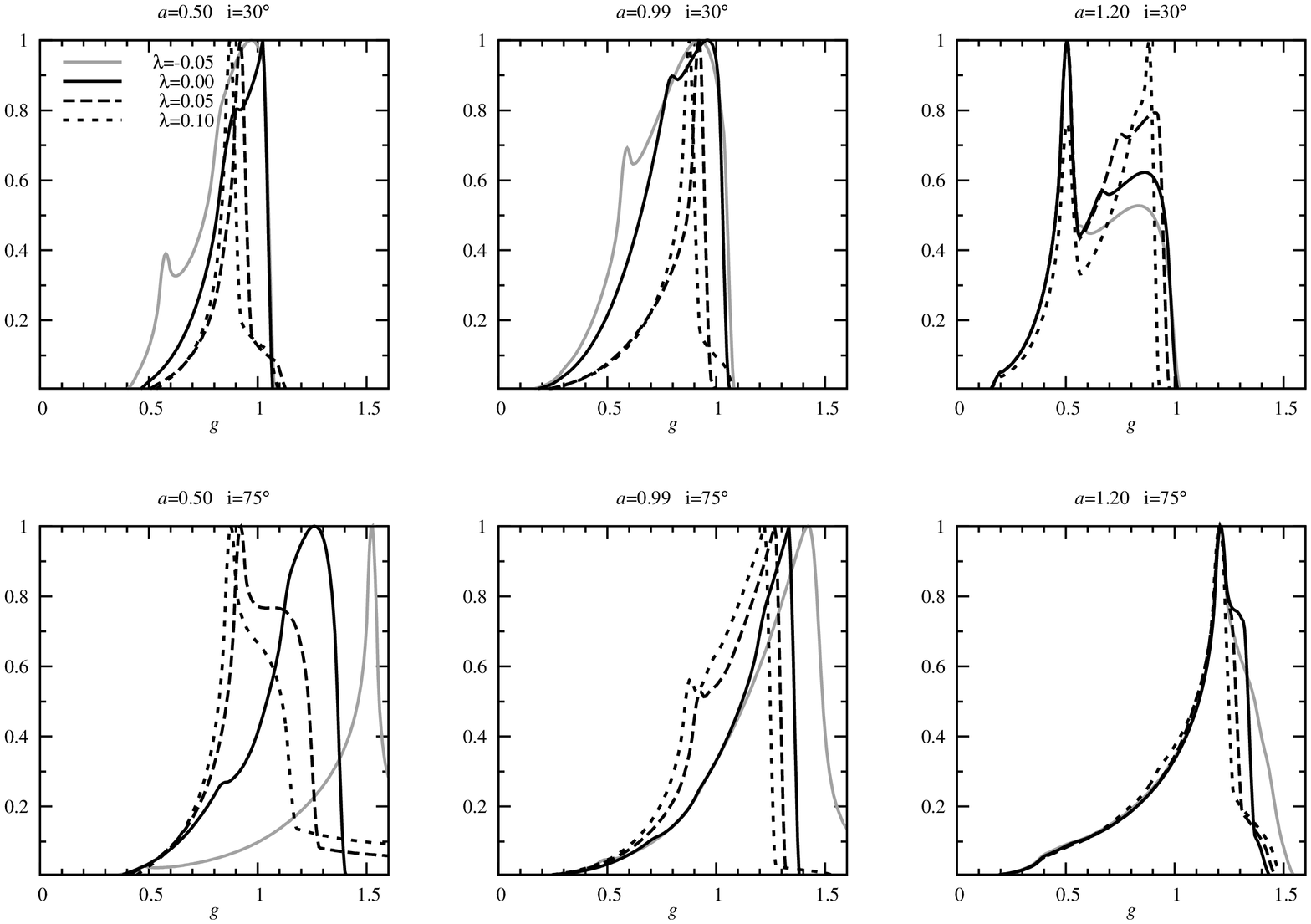}
  \caption{Emission line profiles of accretion discs formed around black holes ($a=0.50$, $0.99$) and naked singularities ($a=1.2$) for two inclinations: $i=30\degr$ ({\bf top} panels) and $i=75\degr$ ({\bf bottom} panels).  The solid lines represent the non-magnetised case. Long-dashed (short-dashed) lines represent uniform magnetic field configurations with $\lambda_{\rm U}=0.05$ ($\lambda_{\rm U}=0.1$). The grey solid line corresponds to $\lambda_{\rm U}=-0.05$. The emissivity index is fixed to $p=3$.}
  
  \label{plot-unif}
  
  \includegraphics[height=0.9\textwidth,angle=-90]{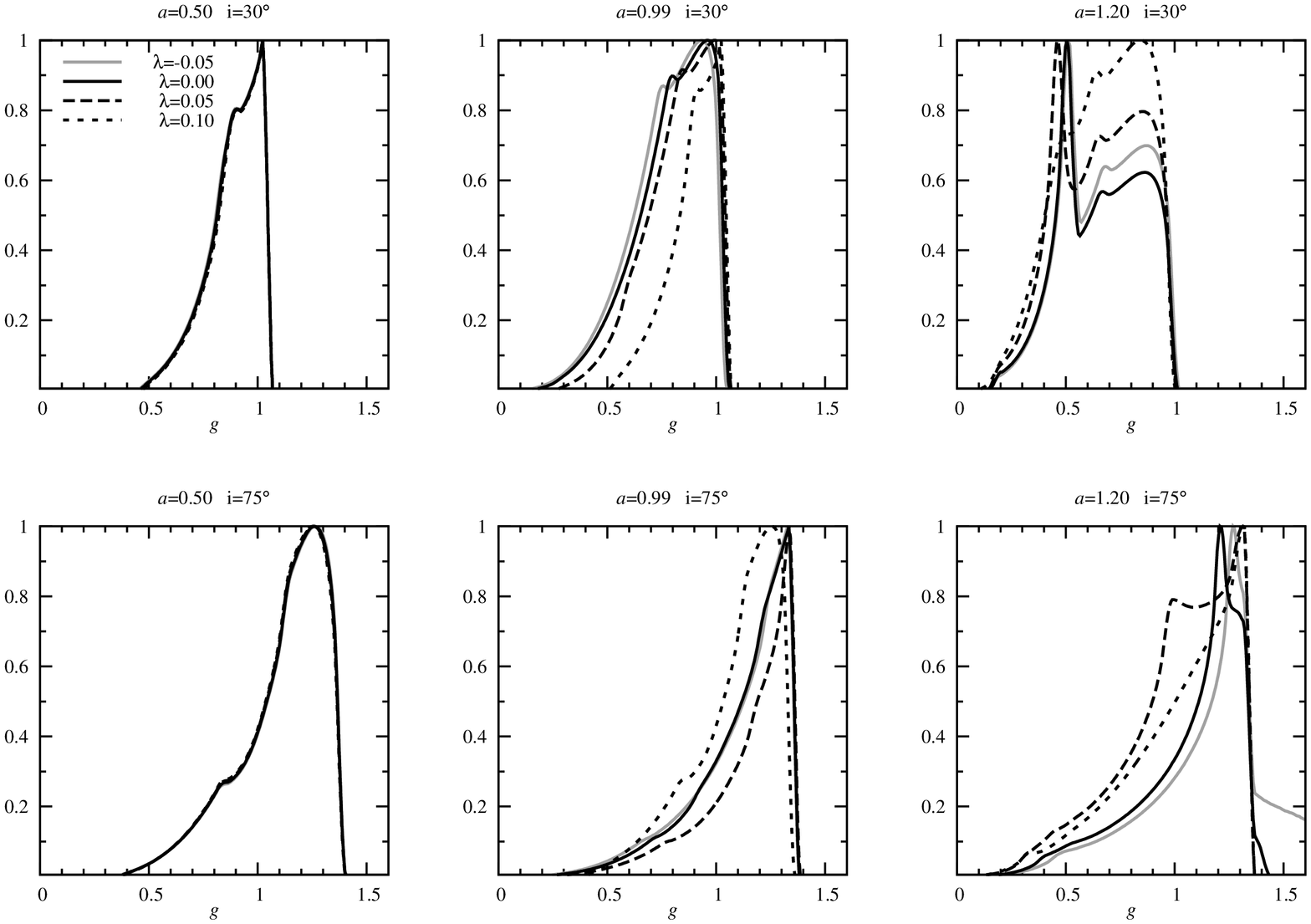}
  \caption{Same as Fig.\ref{plot-unif} for dipolar magnetic field configurations.}
  \label{plot-dip}
\end{figure*}

To analyse possible differences in the observed line profiles arising from different emissivity indexes, in Fig.~\ref{plot1} we present our results for non-magnetised ($\lambda=0$) normalised relativistic line profiles assuming $p=2$ (top panels) and $p=3$ (bottom panels), for fixed values of the spin parameter $a=0$, $0.99$, $1.089$ and $2$, and three different inclination angles $i=0\degr$, $30\degr$ and $75\degr$. The plot shows that, as expected, emission line profiles become harder for higher inclinations. For black holes, the line profiles for low
inclinations look narrower and simpler than those corresponding to naked singularities. For naked singularities, the line profiles show a multi-peak structure because the secondary image of the disc comes from inside the inner region (see right panel of Fig.~\ref{dos-discos}). Our results agree with those found by \citet{schee.stuchlik} for black holes and superspinars. Generally speaking, $p=3$ line profiles are broader than $p=2$ profiles, especially in the red wings. Below we assume $p=3$.

Now we focus on the effects on emission line profiles when an external magnetic field is included. First, in Fig.~\ref{plot-unif} we plot our results for the uniform magnetic field for three different values of $\lambda_{\rm U}=0.05$ and $0.1$, using dashed lines, and $\lambda_{\rm U}=-0.05$ using  solid grey lines. As a reference, we plot the non-magnetised case with solid lines. In the top (bottom) panels we plot the line profiles obtained assuming $i=30\degr$ ($i=75\degr$) and $a=0.5$, $0.99$ and $1.2$. For low values of the spin parameter, $a$, accretion discs extend to outer regions where the effects of the uniform magnetic field configuration dominate the trajectories, and thus, the emission line profiles are strongly affected. In contrast, as $a$ grows and the discs shrink, the effects become less relevant, but still noticeable for these values of $\lambda_{\rm U}$. For $a<1$ and $\lambda>0$, the peak energy of the emission line profiles dims as $\lambda$ grows, while for $\lambda<0$ the opposite occurs. For naked singularities, peak energy and red wings of emission line profiles are only weakly affected.

Secondly, in Fig.~\ref{plot-dip} we present our results for the dipolar magnetic field for the same parameters. In this case, in contrast to the uniform field, the dipolar magnetic field dominates in the inner regions of the discs and thus the main changes in the shape of the line profiles occur for values of $a \sim 1$, as $r_{\rm isco} \rightarrow 1$. This explains why line profiles of magnetised accretion discs for $a=0.5$ are almost indistinguishable from the non-magnetised case, even for $\lambda < 0$. As $a$ grows, this characteristic changes radically, and emission line profiles become more sensitive to the value of $\lambda_{\rm D}$. For extremely fast spinning black holes, the effect of the dipolar magnetic filed on the line profile dominates in the red wings, while for naked singularities, the blue-shifted peak becomes brighter.

\section{Discussion} \label{conclusions}

Using a perturbative approach, we were able to analytically reproduce previously reported numerical results for the radii of the most relevant circular orbits of charged particles orbiting a Kerr black hole with an external uniform and/or dipolar magnetic field, also extending these results to the super-spinning Kerr naked singularity case. Based on these results, and combining them with a public ray-tracing code \citep{ynogk}, we developed a numerical model capable of calculating synthetic energy spectra of the thermal radiation emitted by a relativistic accretion disc formed in a general Kerr background spacetime as well as emission line profiles originating at the surface of the disc, as seen by an arbitrary distant observer. We studied non-magnetised accretion discs, recovering the well-known results from \citet{pyt}, and extended these results by analysing the effects caused in thermal energy spectra and emission line profiles by introducing an external uniform or dipolar magnetic field. 

Uniformly magnetised accretion discs are more affected for slowly spinning black holes and highly spinning naked singularities for which the innermost stable orbit is far from the compact object, and the disc extends to regions where the magnetic field strongly affects the circular trajectories. As the dipolar magnetic field dominates the circular trajectories close to the compact object, the properties of accretion discs are in contrast more affected for nearly extremal black holes and/or naked singularities. Although the effects caused by introducing an external magnetic field on the thermal energy spectra of accretion discs seem hard to distinguish from other uncertainties such as the colour factor, the size of the disc, spin, and inclination, relativistic iron line profiles are heavily affected, and thus we suggest that simultaneous fitting of the thermal spectra and emission line profiles by means of our model can become a useful tool for estimating the presence, intensity, and geometry of the global magnetic field structure of accretion discs around compact objects. 

To compare our theoretical results for the thermal energy spectra of accretion discs with archival observational data of \mbox{X-ray} binaries, we must take into account the \mbox{X-ray} absorption produced by the interstellar medium. Interstellar absorption depends on the cross sections of nuclei and is parameterised by the hydrogen column density, $N_{\rm H}$ \citep{wilms.etal}. To estimate interstellar absorption, we incorporated the analytical fit provided by \citet{int-absX}, for which the cross section is given by $10^{24}\sigma (E) [{\rm cm^2}] = (243.88 - 16.594 E) E^{-2.348}$ for \mbox{X-ray} photons in the $0.532$~keV~$< E < 7.111$~keV energy range.

The black-hole candidate \mbox{Cygnus X-1} is one of the most frequently studied \mbox{X-ray} binaries. Its mass, spin parameter, distance, inclination angle, and mass accretion rate are known with high accuracy. Thus, as an illustrative example, we applied our model to fit \mbox{X-ray} data of this source in the soft state, when the disc emission dominates the \mbox{X-ray} spectra.

\begin{figure}
\centering
\includegraphics[height=0.45\textwidth,angle=-90]{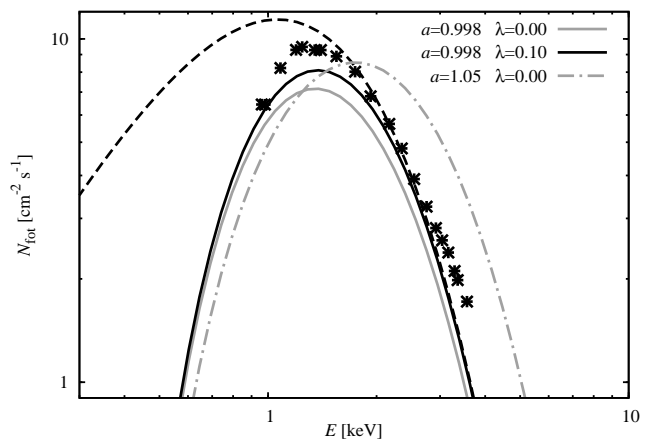}
\caption{\mbox{X-ray} spectra of \mbox{Cygnus X-1}. Crosses represent observational data from \citet{cygX1}. Continuous (dot-dashed) lines correspond to black holes (naked singularities) with $a=0.998$ ($a=1.05$). Black (grey) lines represent uniformly (non-) magnetised accretion discs with $\lambda=0.1$ ($\lambda=0$). The hydrogen column density is set to $N_{\rm H} = 0{.}25 \times 10^{22}$~cm$^{-2}$. As an illustration, the dashed line represents unabsorbed thermal spectra for $N_{\rm H}=0$.}
\label{X1}
\end{figure}

In Fig.~\ref{X1} we present the comparison of our results and the number of observed photons, $N_{\rm fot}$, of \mbox{Cygnus X-1} in the $0.3$~keV~$< E < 10$~keV energy range taken from \citet{cygX1}. The best fit to the data is achieved for a value of the spin parameter ${a=0{.}998}$ (grey solid line). Moreover, an improvement of the fitting is obtained if we add a uniform magnetic field with ${\lambda_{\rm U} = 0.1}$ (black solid line). For this fit, the rest of the parameters, ${f_{\rm col} = 1.8}$, ${\dot{M}M^{-2}}$, ${i}$ and $N_{\rm H}$, are the accepted values for \mbox{Cygnus X-1} \citep{cygX1}. To determine whether a Kerr naked singularity can also model the thermal emission of \mbox{Cygnus X-1}, we present in the same plot the thermal spectra expected for ${a=1.05}$ (dot-dashed line). Although the peak intensity depends on the distance, the observed peak energy for a naked singularity is too hard to fit the observational data of this source, which excludes the possibility of explaining the observed emission with a naked singularity under this accretion-disc model, agreeing with the weak CCC. To show the intensity of the interstellar absorption effects on the observed energy spectra, we present the unabsorbed thermal spectra assuming $N_{\rm H}=0$ as a dashed line.

Answering the question of whether the cosmic censorship conjecture is valid or not is one of the main problems that general relativistic and high-energy astrophysics communities face. The huge technical difficulties of the quest of solving this problem makes astrophysical studies of compact objects highly important because they can become a unique tool for testing the validity of the weak cosmic censorship conjecture. In this context, the development of theoretical tools that can be used to determine the nature of the compact object in an \mbox{X-ray} binary system has been an area of great activity during the last decades. Being able to estimate the nature and strength of the magnetic field in the close region to the compact object present in these types of binaries is of highly interesting and important for understanding the complex scenario in which accretion discs form. As most of the estimates for the magnetic field come by means of properties of the non-thermal radiation produced far away from the disc, that is, using polarimetric $\gamma$-ray observations \citep{laurent,vieyro} and Faraday rotation measurements \citep[see, for example,][]{FRM,FRM2}, studying the effects caused by an external global magnetic field on the thermal emission of accretion discs becomes relevant for these purposes.

Although the most appropiate framework in which to model a magnetised accretion disc in a Kerr background is to study the fully non-linear problem under the magnetohydrodynamic approximation, this quest is beyond the scope of our work. It is important to note that in our model we assumed three fundamental hypotheses: the matter that constitutes the accretion disc does not alter the background spacetime geometry, the magnetic field is not consistently generated by the matter forming the accretion disc, and this matter is a very low-conductivity plasma where the currents are due to bulk motion of the fluid alone. The second assumption is what we call the weak-coupling regime in which matter trajectories are affected by the external magnetic field, but not the other way round. Since our third hypothesis is not completely justified, a more detailed study of this assumption is needed. Although our hypotheses are a strong simplification of the actual physical picture, it is worthwhile to do this because the focus of our work was not the reason or manner in which the magnetic field is generated, but to analytically obtain the observable effects induced by its presence, which we compared with observational data to obtain insight into the nature of these phenomena.

Through this work, we present a tool that can be used to estimate the presence of a magnetic field in the neighbourhood regions of an accretion disc formed around a general Kerr background spacetime. Systematic fitting of thermal spectra and emission line profiles from known \mbox{X-ray} binaries is beyond the scope of this paper. For this purpose, we plan to make public the whole set of results obtained with our model to make our results available to the whole community.

\begin{acknowledgements}
We are grateful to H\'ector Vucetich for valuable suggestions and comments during early stages of this work and to the anonymous referee for constructive input that helped us to improve our manuscript. IFRS acknowledges support from Universidad Nacional de La Plata.
\end{acknowledgements}

\end{document}